# Development of Doped MgB$_2$ Wires and Tapes for Practical Applications

Zhaoshun Gao, Yanwei Ma, Dongliang Wang, Xianping Zhang

*Abstract*—A review of current developments in the study of chemical doping effect on the superconducting properties of MgB$_2$ wires and tapes is presented, based on the known literature data and our own results. The critical current density of MgB$_2$ can be improved through various kinds of dopants. Among these dopants, doping with carbon-containing materials seems to be the most effective way to improve the $J_c$ performance. The doping effect of carbon in different forms and carbon-based compounds such as SiC, nano-C, metal carbides, as well as aromatic hydrocarbon and carbohydrate on the $J_c$-$B$ characteristics of MgB$_2$ was discussed in detail. The C can be incorporated into the MgB$_2$ crystal lattice by replacing boron, and thus $B_{c2}$ is significantly enhanced due to selective tuning of impurity scattering of the $\pi$ and $\sigma$ bands in the two-band MgB$_2$. Besides the efforts of increasing $B_{c2}$ by carbon doping, the fine grain size and nano-size inclusions caused by doping would create many flux pinning centres improving the $J_c$-$B$ property of MgB$_2$. Based on these considerations, we suggested some principles for the selection of dopants.

## I. INTRODUCTION

MgB$_2$ is a promising superconductor for practical applications in the field of superconducting magnets for MRI, due to its rather high $T_c$, relatively low material costs and weak-link-free behavior[1]. The development of MgB$_2$ tapes and wires was much faster than for many others HTS and LTS materials and commercial wires and tapes became available only a few years after the discovery of MgB$_2$.

From the view point of applications, the critical current density $J_c$ in high magnetic fields is crucially important. However, the pristine MgB$_2$ always shows lower $J_c$ values because of low upper critical field ($B_{c2}$) and poor flux pinning. In order to improve $J_c$–$B$ properties, a number of experimental techniques, including chemical doping [2, 3], irradiation [4], magnetic field annealing [5, 6], and ball-milling methods [7], have been attempted. Compared to other methods attempted, chemical doping with carbon-containing materials was thought as the convenient and effective way to enhance the $J_c$-$B$ properties of MgB$_2$. Therefore, effects of such additives are being actively investigated, and some effective dopants have been suggested [2, 3, 8-15]. The additions of SiC, nano-C, carbon nanotubes, B$_4$C, as well as aromatic hydrocarbon and carbohydrate are effective to improve the $J_c$-$B$ characteristics of MgB$_2$ [2, 3, 10-15]. When MgB$_2$ is doped with these materials, C substitutes for B and introduces electron scattering and impurity scattering which reduce the mean free path and the coherence length ξ, hence increasing $B_{c2}$ [16]. Increases in $B_{c2}$ can increase the pinning strengths of already existing pins. At the same time, they are also effective in the enhancement of pinning strength by introducing lattice distortion and more grain boundaries.

Here we present a review of the recent results in the doping of MgB$_2$ tapes and wires using different types of carbon sources and some promising processes combined with carbon doping. Based on these results, we suggest some principles for the selection of dopants and give a perspective for the future development of this superconductor.

## II. CARBON DOPING

The effects of carbon doping on superconductivity in MgB$_2$ compound has been extensively studied. The $J_c$-$B$ performance can be greatly improvement by doping with various different carbon sources, such as nanocarbon [11, 17], diamond [18], graphite [19], and carbon nanotubes [12].

The authors' group has systematically studied the nano-C doping effect on the critical temperature, $J_c$-$B$ property and $B_{c2}$ of MgB$_2$ tapes [11, 20-21]. Soon thereafter, our results have been confirmed by many groups worldwide [22-24]. The addition of nano-C causes substitution of boron by carbon, which decreases the critical temperature and increases the upper critical field as well as the current density in high magnetic fields. Fig. 1 shows the transition temperature curves of tapes with different C doping levels determined by susceptibility measurements. The $T_c$ decreases monotonically with increasing nano-C doping level. The depression of $T_c$ is caused by the carbon substitution for B. On the other hand, the carbon substitution for B was found to enhance $J_c$ in magnetic fields. Figure 2 shows the transport $J_c$ at 4.2 K in magnetic fields for Fe-sheathed MgB$_2$ tapes with various amounts of nano-C doping from 0 to 10 at% that were heat-treated at 650 °C. It can be seen that, in measuring fields of up to 14 T. all the doped tapes exhibited superior $J_c$ values compared to the pure tape. The highest $J_c$ value of the Fe-sheathed tapes was achieved by the 5 % nano-C addition; further increasing C

Manuscript received October 28, 2009. This work is partially supported by the National Natural Science Foundation of China under Grant No. 50802093, National '973' Program (Grant No. 2006CB601004) and International Cooperation Project between China and Japan (Project No. 2008ICC-PROJECT-05).

Zhaoshun Gao, Yanwei Ma, Dongliang Wang and Xianping Zhang are with the Applied Superconductivity Lab., Institute of Electrical Engineering, Chinese Academy of Sciences, P. O. box 2703, Beijing 100190, China (e-mail: gaozs@mail.iee.ac.cn, ywma@mail.iee.ac.cn).

doping ratio caused a reduction of $J_c$ in magnetic fields.

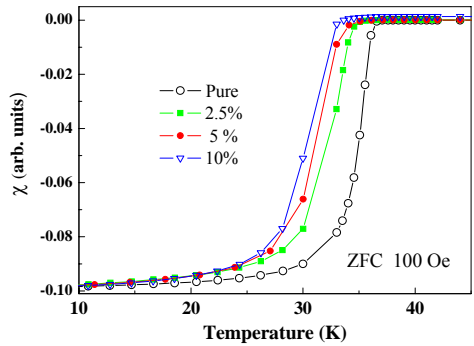

Fig. 1. Normalized magnetic susceptibility versus temperature for the samples with different doping level [11].

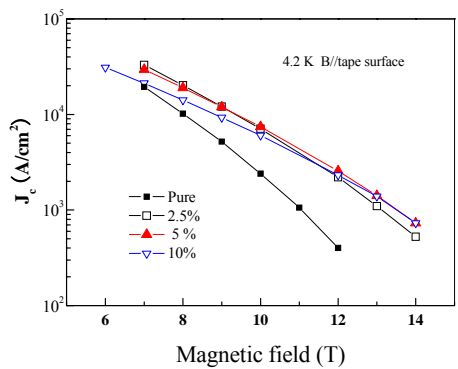

Fig. 2. Critical current densities at 4.2 K versus magnetic field in MgB$_2$ tapes with nano-C doping level from 0 to 10 at.%, which were heat-treated at 650 °C.

Fig. 3 shows the $J_c$ at 4.2 K and 10 T for 5% C-doped tapes that were sintered at different temperatures ranging from 600 to 950 °C. From Fig. 3, we immediately notice that the sintering temperature has a significant effect on the $J_c$-$B$ performance. The $J_c$ values of doped samples increased systematically with increasing sintering temperatures. The tape sintered at 950 °C exhibited the highest $J_c$ values with excellent $J_c$-$B$ performance compared to all other samples: at 4.2 K, the transport $J_c$ reached $2.11 \times 10^4$ A/cm$^2$ at 10 T. Higher annealing temperatures promote the reaction of C substitution for B [22], thus enhancing flux pinning and improving the high-field $J_c$.

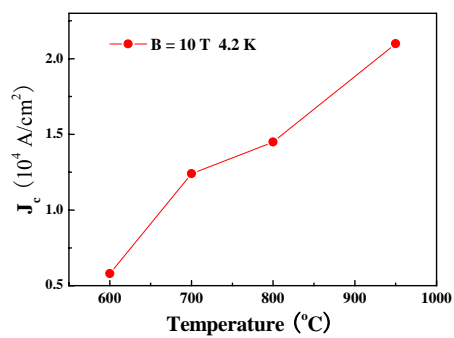

Fig. 3. Transport $J_c$ at 4.2 K in magnetic fields up to 18 T for 5 at.% C-doped MgB$_2$ tapes sintered at various temperatures.

So far, the only element well known to increase $B_{c2}$ is carbon. The significant improvements of $B_{c2}$ in carbon doped samples have been reported by several groups [25–28]. Figure 4 shows the temperature dependence of $B_{c2}$ and $B_{irr}$ for the pure MgB$_2$ tapes and nano-C doped samples, where the $B_{c2}$ and $B_{irr}$ obtained from the 10% and 90% values of the normal-state resistance $\rho$. Clearly, as a result of nano-C doping, the $B_{c2}$ ($B_{irr}$)-$T$ curve became steeper, indicating an improved $B_{c2}$ and $B_{irr}$. For instance, at 20 K, the $B_{irr}$ achieved 9 T for C-doped tapes, which was comparable to the $B_{c2}$ at 4.2 K of NbTi conductors [29].

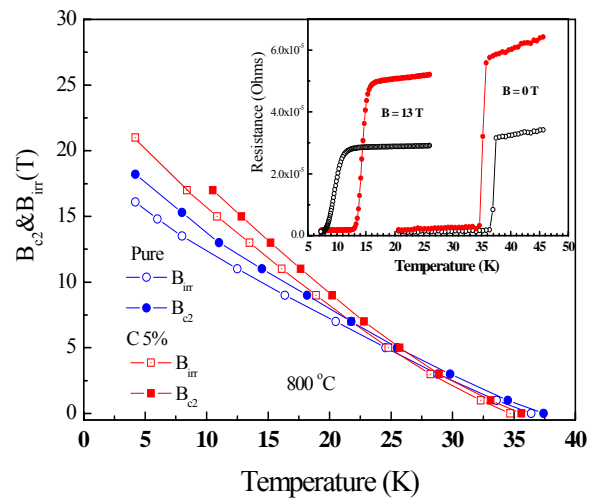

Fig. 4. $B_{c2}$ and $B_{irr}$ values as a function of the temperature for the pure and 5 at.% C-doped MgB$_2$ tapes.

The type of carbon sources is an important parameter determining the effect of doping on the $J_c$-$B$ performance. Figure 5 shows the transport $J_c$ at 4.2 K in magnetic fields for MgB$_2$ tapes with various carbon sources studied by our group. It should be noted that the $J_c$ value of tapes doped with activated carbon was lower than for samples doped with other carbon sources. The average particle size of activated carbon is over 200 nm, much larger than that of nano-C. So the degree of reacting activated carbon particles with MgB$_2$ is much lower than in the case of nano-C particles. Thus too few carbon atoms are incorporated into the MgB$_2$ lattice during its formation. Also, these large particles existing in MgB$_2$ matrix will block the superconducting current flow, and thus decrease the MgB$_2$ superconducting connectivity. However, the $J_c$-$B$ properties for hollow carbon spheres (HCS) and C$_{60}$ doped tapes are better than for nano-C-doped samples. Compared to nano-C dopants, HCS and C$_{60}$ have very special geometrical configuration. They can be broken into nanoflakes and particles during the process. These nanocarbon flakes and particles would easily incorporate into the crystal lattice of MgB$_2$ and substituted into B sites via the reaction.

Haßler et al. [33] have reported the highest $J_c$ (10$^4$A/cm$^2$ at 16.4 T and 4.2 K) at high magnetic fields by high-energy ball milling technique with nanocarbon alloying. This excellent performance can be attributed to the nanocrystalline grain size with a high amount of grain boundaries, and the effectivity of carbon substitution by mechanical alloying [34, 35]. The grain boundaries contribute to better flux pinning and the carbon doping greatly improves $B_{c2}$ by impurity scattering. The results demonstrated that nano-C doping combined with high energy ball milling techniques is an effective method for further enhancing the current capacity of $MgB_2$.

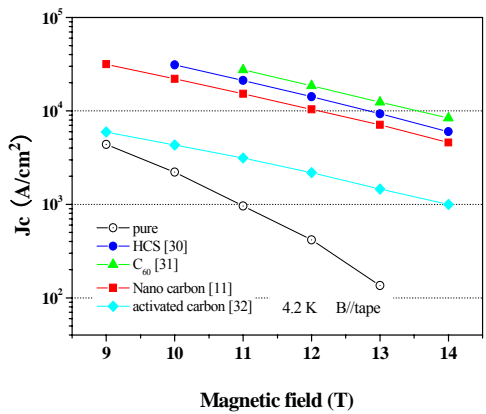

Fig. 5. The transport $J_c$ at 4.2 K in magnetic fields for $MgB_2$ tapes with various carbon sources studied by our group.

### III. SILICON CARBIDE DOPING

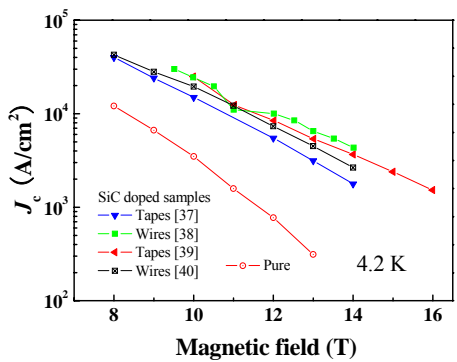

Fig. 6. A comparison of $J_c$ for SiC doped samples from various groups.

The significant enhancement of $J_c$, $B_{irr}$, and $B_{c2}$ in $MgB_2$ by nano-SiC doping was first shown by Dou's group [2]. A high $B_{irr}$ of 29 T and $B_{c2}$ of 37 T have been achieved for nano-SiC doped $MgB_2$ wires [3, 36]. Fig. 6 shows a comparison of transport $J_c$ at 4.2 K for SiC doped samples by the authors' group [37], Dou et al. [38], Matsumoto et al. [39], and Tomsic et al. [40]. The in-field $J_c$ for the nano-SiC doped samples increased by more than one order of magnitude, compared with the undoped samples. Dou et al. have proposed a dual reaction model to explain the enhancement in electromagnetic properties by SiC doping [38]. In this model, fresh reactive C is released from the SiC at low temperature (600 °C), when SiC reacts with Mg to form $Mg_2Si$. This reactive C can effectively substitute into B sites, which leads to the enhancement of the $B_{c2}$. In addition, the by-products, such as highly dispersed fine particles, including $Mg_2Si$, and the excess nanosize C can be embedded into the $MgB_2$ matrix and act as flux pinning centers.

### IV. ORGANIC COMPOUNDS DOPING

Yamada et al. have reported that aromatic hydrocarbon addition to $MgB_2$ can enhance the flux pinning in $MgB_2$ tapes [13]. However, this organic solvent is very volatile at ambient pressure and it is difficult to control the doping level. Later, Kim et al. have reported that carbohydrate-doped $MgB_2$ bulks exhibited a positive effect in improving the $J_c$ values at high magnetic field [14]. Our group's recent results [15, 41-42] show that the $J_c$-$B$ performance of $MgB_2$ tapes with carbohydrate dopant can be as good as for nano-SiC and nano-C dopants [2, 11]. Organic material can decompose at relatively low temperature, and generate reactive carbon atoms before the $MgB_2$ phase formation. The reactive carbon atoms and small sized impurities are favorable for good superconducting properties of $MgB_2$. Compared to nano-sized C or SiC, organic material doping can achieve a more uniform dispersion within the $MgB_2$ matrix [43].

We have investigated the effects of several organic compound dopants on the superconducting properties of $MgB_2$ tapes [15, 41-42]. Our results demonstrated that both the $J_c$ and flux pinning ability of $MgB_2$ tapes are significantly improved through these dopants. Compared to the undoped samples, all doped tapes showed an enhancement of $J_c$ values in high-field region by more than an order of magnitude, as shown in Fig. 7. For example, at 4.2 K and in a field of 12 T, the $J_c$ value of the 10 wt% maleic anhydride doped sample reached $9.2\times10^3$ A/cm$^2$, more than 23-fold improvement compared to the pure samples. These $J_c$ values of the tapes investigated in this work are comparable to the best nano SiC doped tapes reported so far [38], and highlight the importance of organic compounds doping for enhancing the $J_c$ of $MgB_2$ superconductors.

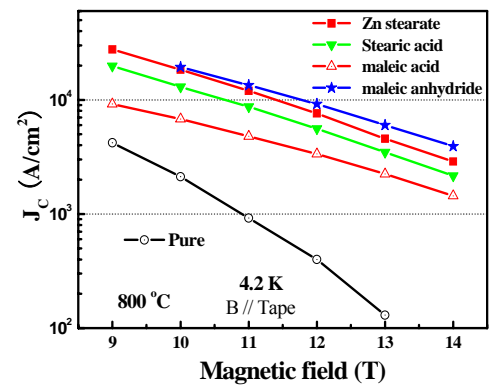

Fig. 7. $J_c$-$B$ properties of tapes doped by pure and organic compounds.

Fig. 8 shows the typical TEM micrographs for the Zn stearate and stearic acid doped tapes. When an organic compound is heated at above the decomposition temperature, the highly reactive C released from decomposition can effectively substitute into B sites. At the same time, C substitution causes reduction in the grain size, and hence enhances the grain boundary pinning. Moreover, the TEM examination revealed that there are a number of impurity phases in the form of nano-size inclusions in doped samples. Therefore, a combination of C substitution-induced $B_{c2}$ enhancement as well as the strong flux pinning centers are responsible for the superior $J_c$-$B$ performance of organic compounds doped samples.

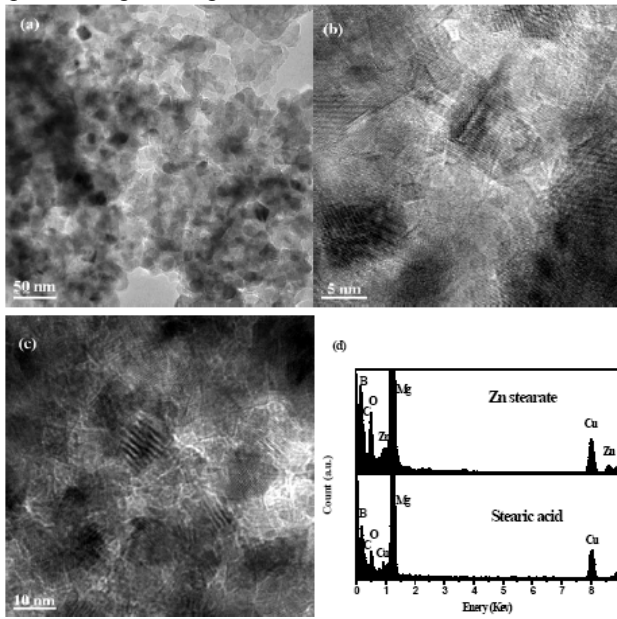

Fig. 8. The TEM images of the Zn stearate and stearic acid doped samples.

The authors' group also found that the oxygen amount in carbohydrate additives strongly affect $J_c$-$B$ performance of MgB$_2$ [42]. The transport $J_c$ and the connectivity were observed to decrease as oxygen content in dopants increased. The higher MgO content and porosity formed during the reaction in high oxygen doped samples is responsible for their low $J_c$ values. As a result, the decrease of oxygen content in dopants would be a key point to further improve the transport $J_c$ of MgB$_2$.

Recently, Flukiger's goup found that the $J_c$ and $B_{irr}$ in C$_4$H$_6$O$_5$ alloyed MgB$_2$ wires were significantly enhanced by cold high pressure densification (CHPD) [44]. C$_4$H$_6$O$_5$ alloyed MgB$_2$ square wires after conventional processing were submitted to cold high pressure densification (CHPD), resulting in the highest $J_c$ values reported so far for nearly isotropic *in situ* wires: $B(10^4)^\perp = 12.5$ T and $B(10^4)^\| = 12.7$ T at 4.2 K. Here, $B(10^4)$ means the field at which $J_c = 1 \times 10^4$ A/cm$^2$. The results demonstrate that organic material doping combined with CHPD techniques is one of promising methods for the enhancement of current capacity of MgB$_2$.

## V. METAL CARBIDE DOPING

The effect of metal carbides like ZrC [45], NbC [46], TaC and MoC$_2$ [47] has been studied in our group. We found that these carbides show little effect on $J_c$, because they are stable in contact with Mg and B even at high temperature; thus no fresh reactive C is available for incorporating into the MgB$_2$ lattice during its formation.

## VI. SIMULTANEOUS ADDITIVES

Yamada *et al.* found that the additive combination of ethyltoluene + SiC were very effective for the improvement of $J_c$-$B$ properties of MgB$_2$ tapes than are either addition alone [48]. The highest $J_c$ values at 4.2 K reached $1.4 \times 10^4$ A/cm$^2$ in 12 T for tapes with added ethyltoluene and SiC. This can be attributed to a cumulative effect on $J_c$ - one comes from the addition of ethyltoluene and the other comes from the carbon substitution for boron by the SiC addition.

The substitution of boron by carbon is known to enhance the impurity scattering and thus of the critical field. In addition, the pinning behavior is expected to be improved by nanosize precipitations. Since the two mechanisms are independent, their effect on $J_c$ is expected to be cumulative [49, 50].

Flukiger's group has reported on the effects of additive combinations B$_4$C+SiC and B$_4$C+LaB$_6$. For both additive combinations, $J_c$(20 K) at high fields is markedly higher than for SiC additives, in contrast to the data at 4.2 K. Mikheenko *et al.* [51] used a combination of Dy$_2$O$_3$ for pinning, together with B$_4$C for doping and successfully improved the $J_c$-$B$ performance of MgB$_2$ in the intermediate field regime (2–5 T) at 20 K. The optimum amount of Dy$_2$O$_3$ and B$_4$C additions is 0.5 wt % of Dy$_2$O$_3$ and 0.04 wt % of B$_4$C, yielding a $J_c$ (20 K) of $10^5$ A/cm$^2$ at 2.7 T. These results demonstrate that simultaneous additives are promising for applications at 20 K.

Recently, our group realized the simultaneous introduction of "scattering + pinning" using a single dopant of organic rare earth salt. Figure 9 shows the XRD results. Rare earth borides and MgO were detected as impurity phases in doped samples. It can be seen that the position of (110) peak slightly shifts to higher angles due to the organic rare-earth salt addition, meaning a decrease in the a-axis lattice parameter. It indicates that carbon has been doped into the B sites in the crystal lattice.

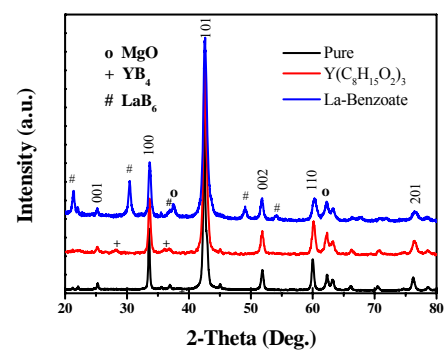

Fig. 9. XRD results for the samples doped with organic rare-earth salts.

Fig. 10 shows the transport $J_c$ at 4.2 K in magnetic fields for two organic rare earth salts doped samples. It is evident that both doped samples exhibited superior $J_c$ values compared to the pure one. For example, at 4.2 K and in a field of 12 T, the $J_c$ value for La-Benzoate doped tape reached $1.84 \times 10^4$ A/cm$^2$, 23 times higher than that of undoped tape.

When organic rare-earth salt is decomposed at low temperature, the highly reactive C can effectively substitute into B sites. Simultaneously, the rare-earth boride inclusions with nano-sizes resulting from the reaction also can act as effective flux pinning centers and hence improve the $J_c$ performance of organic rare-earth salt doped samples.

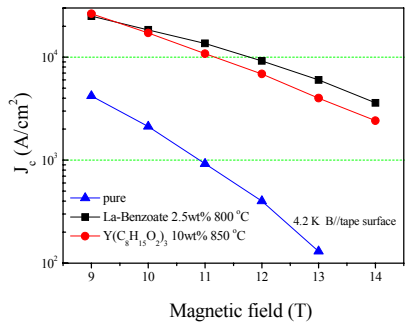

Fig. 10. $J_c$-$B$ properties of pure and organic rare-earth salts doped tapes.

## VII. Discussion

In the past several years, various kinds of dopants have been introduced into $MgB_2$ [8, 52-53]. Fig. 11 shows the doping effects on the $J_c$-$B$ properties of $MgB_2$ with various kinds of dopants studied in our group. Based on the results mentioned above, we suggest some guiding principles for the selection of dopants.

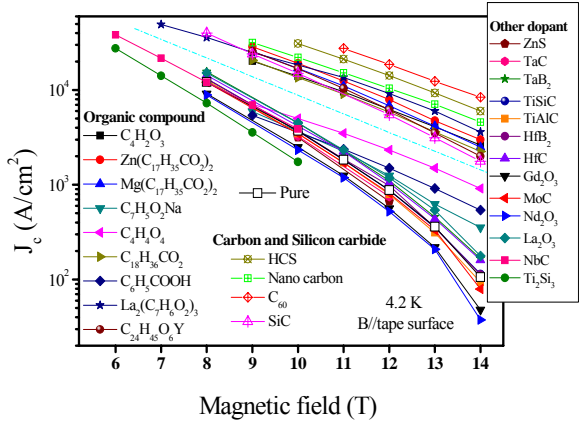

Fig. 11. The transport $J_c$ at 4.2 K in magnetic fields for $MgB_2$ tapes with various kinds of dopants studied in our group [47].

1. Dopants should contain carbon. Carbon is the only element confirmed to be able to remarkably enhance $B_{c2}$ of $MgB_2$.
2. In selecting dopants, reaction chemistry is an important consideration. Carbon-containing dopants with high reactivity decompose or react at low temperatures, releasing fresh C which can substitute for B efficiently, resulting in small grains.
3. Proper amount of secondary phases induced by carbon sources may enhance the vortex pinning, but high amount of impurities strongly degrade the connectivity if present between $MgB_2$ grains, as seen in O-contaminated $MgB_2$ [42].

The doping directly influences the $B_{c2}$, flux pinning and connectivity of $MgB_2$. We should carefully tune the balance between them and optimize the doping effects.

Many dopants can significantly enhance the $B_{c2}$ and $J_c$ of $MgB_2$, but often at the expense of homogeneity and connectivity. Furthermore, the porosity of *in situ* made $MgB_2$ greatly reduces the super current path. More research and new techniques are needed to increase homogeneity and connectivity of the $MgB_2$.

## VIII. Conclusion

The experimental results show that both the $B_{c2}$ and $J_c$ of $MgB_2$ wires and tapes doped with carbon sources are significantly enhanced. The performance of $MgB_2$ is expected to further increase in the coming years through improved flux pinning, homogeneity and connectivity. In particular, the excellent $J_c$-$B$ performance of $MgB_2$ wires and tapes was achieved through C-based doping combined with CHPD or high energy ball milling techniques. This result demonstrates that $MgB_2$ has potential for use in high magnetic fields and as a strong competitor for the currently used NbTi and $Nb_3Sn$ conductors.

## Acknowledgment

The authors thank Liye Xiao, Xiaohang Li and Liangzhen Lin for their help and useful discussion.